\documentclass{article}

\PassOptionsToPackage{numbers, compress}{natbib}

\usepackage{hyperref}
\usepackage{amsmath,amssymb,amsfonts}
\usepackage{algorithm}
\usepackage{algorithmic}
\usepackage{graphicx}
\usepackage{pifont}
\usepackage{wasysym}
\usepackage[preprint]{neurips_2024}
\usepackage{subfigure}
\usepackage{natbib}
\usepackage{textcomp}
\usepackage{xcolor}
\usepackage{verbatim}
\usepackage[capitalize,noabbrev]{cleveref}
\usepackage[most]{tcolorbox}
\usepackage[utf8]{inputenc}

\newtcolorbox{questionbox}[1][]{
  colback=#1,
  colframe=black,
  boxrule=1pt,
  arc=5pt,
  outer arc=5pt,
  enhanced,
  breakable
}

\definecolor{boxcolor1}{RGB}{230,240,255} 
\definecolor{boxcolor2}{RGB}{255,240,230} 
\definecolor{boxcolor3}{RGB}{230,255,230} 
\definecolor{boxcolor4}{RGB}{255,230,255} 
\definecolor{boxcolor5}{RGB}{255,255,230} 
\definecolor{boxcolor6}{RGB}{230,255,255} 

\usepackage[utf8]{inputenc} 
\usepackage[T1]{fontenc}    
\usepackage{hyperref}       
\usepackage{url}            
\usepackage{booktabs}       
\usepackage{amsfonts}       
\usepackage{nicefrac}       
\usepackage{microtype}      
\usepackage{xcolor}         

\title{Interpreting Multi-band Galaxy Observations with Large Language Model-Based Agents}

\author{%
  Zechang Sun\\
  Department of Astronomy\\
  Tsinghua University\\
  Beijing, China\\
  \texttt{szc22@mails.tsinghua.edu.cn} \\
  \And
  Yuan-Sen Ting \\
  Department of Astronomy \\
  The Ohio State University\\
  Columbus, OH 43210, USA\\
  \texttt{ting.74@osu.edu} \\
  \AND
  Yaobo Liang \\
  Microsoft Research Asia \\
  Beijing, China\\
  \texttt{yaobo.liang@microsoft.com} \\
  \And
  Nan Duan \\
  Microsoft Research Asia \\
  Beijing, China \\
  \texttt{nanduan@microsoft.com} \\
  \And
  Song Huang \\
  Department of Astronomy \\
  Tsinghua University \\
  Beijing, China\\
  \texttt{shuang@mail.tsinghua.edu.cn} \\
  \And
  Zheng Cai \\
  Department of Astronomy \\
  Tsinghua University\\
  Beijing, China\\
  \texttt{zcai@mail.tsinghua.edu.cn}\\
}

\begin{document}

\maketitle

\begin{abstract}
Astronomical research traditionally relies on extensive domain knowledge to interpret observations and narrow down hypotheses. We demonstrate that this process can be emulated using large language model-based agents to accelerate research workflows. We propose \texttt{mephisto}, a multi-agent collaboration framework that mimics human reasoning to interpret multi-band galaxy observations. \texttt{mephisto} interacts with the \texttt{CIGALE} codebase, which includes spectral energy distribution (SED) models to explain observations. In this open-world setting, \texttt{mephisto} learns from its self-play experience, performs tree search, and accumulates knowledge in a dynamically updated base. As a proof of concept, we apply \texttt{mephisto} to the latest data from the James Webb Space Telescope. \texttt{mephisto} attains near-human proficiency in reasoning about galaxies' physical scenarios, even when dealing with a recently discovered population of "Little Red Dot" galaxies. This represents the first demonstration of agentic research in astronomy, advancing towards end-to-end research via LLM agents and potentially expediting astronomical discoveries. \end{abstract}
\section{Introduction}

The advent of deep learning tools and the vast amount of data routinely collected in astronomical surveys—from hundreds of millions of spectra \cite{SDSS2000,4MOST2019,DESIEDR2024} to tens of billions of images \cite{LSST2019,EUCLID2022,HSC2022}—has propelled astronomical research into high gear. However, much of the development in AI-assisted astronomy often focuses on optimizing individual downstream tasks, such as building classifiers and brokers to streamline surveys \cite{BEN2022,STRONGLENSING2022,BRANT2023}, emulating expensive hydrodynamical simulations \cite{CHARDIN2019,SIYUHE2019,YINLI2021}, and advancing statistical inference using generative models as posterior and likelihood surrogates \cite{CRANMER2020,XIAOSHENG2022,SUN2023a}. Despite these advancements, focusing solely on individual downstream tasks limits the potential impact of AI in astronomical research.

Finding interesting astronomical phenomena based on limited and noisy observations often requires researchers to sift through a vast array of possible hypotheses to find viable and rational explanations. As an exhaustive search of all possible hypotheses to match the observations is computationally infeasible, the process of reasoning, filtering, and reflecting remains one of the crowning jewels of modern human innovation \cite{ABDO2009,GILLON2016,SPITLER2016,WANG2021,WELCH2022,MAIOLINO2024}. This process essentially allows researchers to find solutions through "intuition" and experience—capabilities that go beyond optimizing individual downstream tasks.

With the growing trend of adopting large language models to design and conduct surveys in scientific research \cite{YOSH2023,LEI2024,JAB2024}, a key question emerges: Can AI autonomously learn from its own experiences, reflect on failures and successes like humans, and ultimately build up enough knowledge to perform reasoning on physical models against observational data in astronomy without extensive exhaustive search? To explore this, we have developed a multi-agent collaboration framework, \texttt{mephisto}. We focus on a specific astronomical research problem: reasoning about the task of fitting broadband photometric observations, known as Spectral Energy Distributions (SEDs), which can be seen as extremely low-resolution spectra of galaxies.

As a proof of concept, we apply \texttt{mephisto} to the latest data from the James Webb Space Telescope's JADES survey. The unique wavelength coverage and unprecedented spatial resolution of James Webb, the successor to the Hubble Space Telescope, has led to a trove of discoveries, including some of the highest redshift (earliest) galaxies ever observed \cite{COSMOSWEB2023,JADES2023,UNCOVER2024}, as well as a new population of galaxies known as the "Little Red Dots" \cite{PABLO2024,MATT2024,GENTILE2024,BAGGEN2024}. Interpreting the physics of galaxies through multi-band observations is fundamental to modern galaxy physics and cosmology. Key inquiries in galaxy evolution—including dark matter dynamics \cite{COLLINS2009,MANDELBAUM2018,DESJACQUES2018,WECH2018} and the role of black holes \cite{CATTANEO2009,FABIAN2012,ABB2022,WANG2024a}—hinge on accurate interpretation of galaxy SEDs \cite{BRAMMER2008,CHEV2016,CIGALE2019,JOHN2021}. The process of identifying model-data discrepancies, iteratively refining SED models, and developing viable explanations \cite{PACI2023,PABLO2024,BINGJIE2024} requires extensive domain expertise and is time-intensive for researchers. Our work aims to emulate and accelerate this complex process using LLM-based agents.

\section{Method}\label{sec:method}

\begin{figure*}
    \centering
    \includegraphics[scale=0.35]{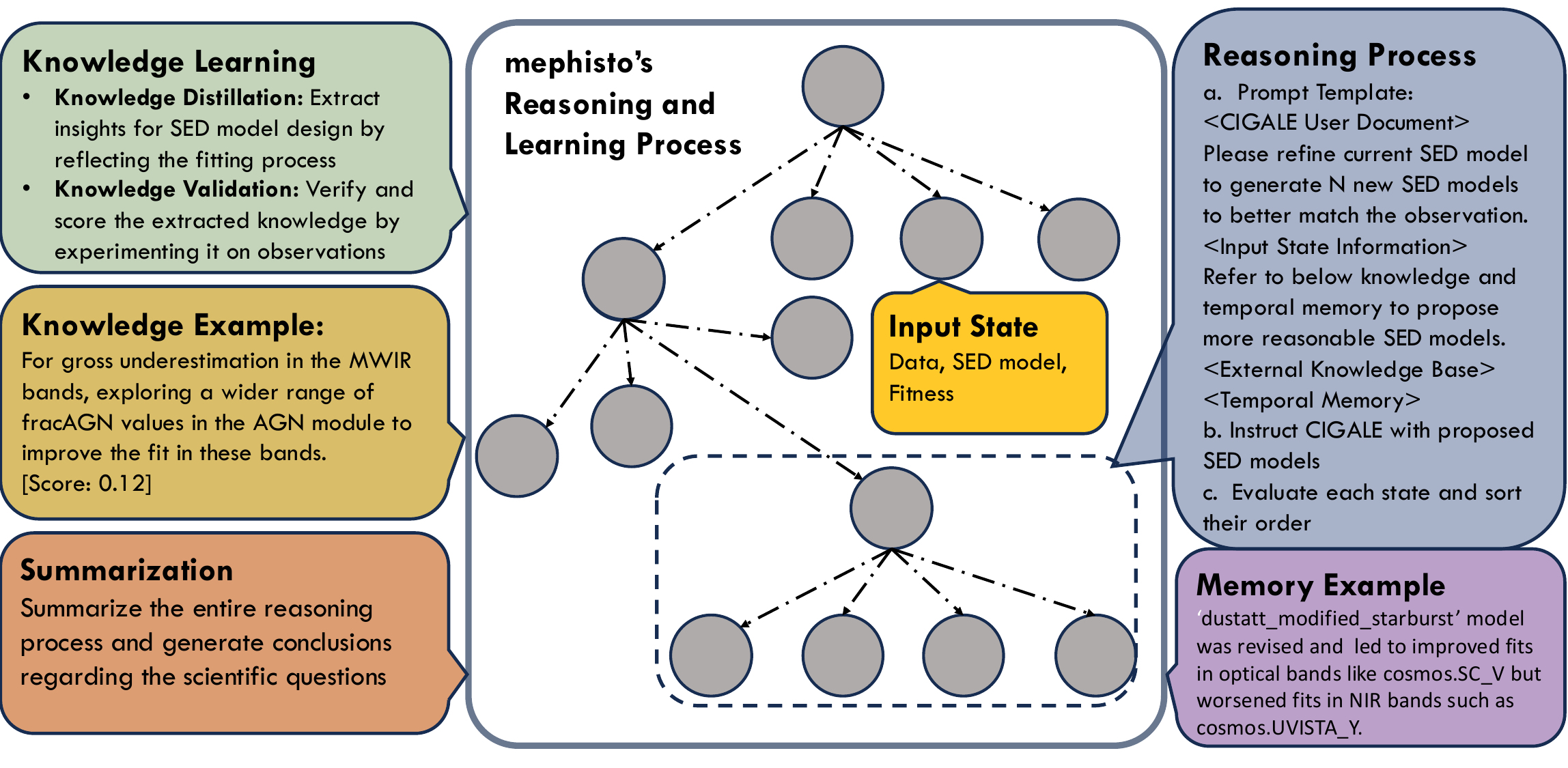}
    \caption{Schematic illustration of \texttt{mephisto}'s process for interpreting multi-band galaxy observations. Starting with a base SED model, \texttt{mephisto} assesses discrepancies between model predictions and observational data. It then generates four hypotheses (model variants) for improvement, which are input into \texttt{CIGALE} for refined fitting. \texttt{mephisto} selects the most promising unexpanded model variant based on fit quality, balancing model complexity with tree search depth. The process incorporates temporal memory to enhance efficiency and diversity of SED model proposals. Outcomes are compiled and distilled into a dynamically updated knowledge base through a learning system that autonomously extracts and validates knowledge. This iterative self-play process mimics human reasoning, allowing \texttt{mephisto} to achieve near-human proficiency in interpreting galaxy SEDs.}
    \label{fig:schema}
\end{figure*}

\texttt{mephisto} is a multi-agent collaboration framework designed to emulate human expert reasoning in fitting SED models. As illustrated in Figure~\ref{fig:schema}, \texttt{mephisto} iteratively interacts with \texttt{CIGALE}\cite{CIGALE2019} to construct increasingly complex SED models. These models incorporate various components including star formation histories \cite{CIGALE2019}, stellar populations \cite{BRUZUAL2003,MARASTON2005}, dust attenuation and emissions \cite{LEITHERER2002,CALZETTI2000,DALE2014}, nebular emissions \cite{FERLAND1998,FERLAND2013}, and active galactic nuclear emissions \cite{FRITZ2006,CAMPS2015}. To enhance model diversity and efficiency, we implement temporal memory and an external knowledge base, which is dynamically updated through knowledge distillation and validation, further improving the framework's performance over time, mimicking the learning process of a human expert.

\paragraph{Input State} \texttt{mephisto} can be viewed as a reinforcement learning process without an explicit reward function, where an LLM agent implicitly evaluates the reward. The framework's goal is to assess the current SED "state" and propose actions to optimize the goodness of fit to JWST data. Formally, the input state $s(d,m,r)$ (in json format, see Appendix~\ref{appendix:input} for details) comprises observation data $d$, described as wavelength-flux tuples due to current limitations in vision language models' ability to interpret scientific plots \cite{LILEI2024,YUE2023}; SED model $m$, which includes physical model names, parameter ranges, priors, and grid sizes; and fitting results $r$, containing $\chi^2$ information and auxiliary data such as the number of well-fitted photometry bands, as determined by the evaluation agent. This input structure, particularly the vague definition of well-fitted bands, mimics human practices. It allows the agent to consider overall S/N and global fit quality, potentially identifying corrupted data. This approach highlights the importance of agentic research in SED studies, where a physically implausible model with optimal $\chi^2$ is often disfavored. The evaluation relies on human-like fuzzy logic and decision, which we aim to emulate, rather than solely on single statistical metric.

\paragraph{Reasoning Process} \texttt{mephisto} guides \texttt{CIGALE} through an exploration of various SED models, iteratively seeking plausible interpretations of observations. At each step, \texttt{mephisto} analyzes the state $s(d,m,r)$, reasons about discrepancies between model predictions and data, and generates $N_b=4$ hypotheses to refine the current SED model. This iterative approach uses a tree structure to facilitate deliberate reasoning \cite{YAO2023,GOT2023,AOT2023,CHEN2022} and mirrors human reasoning processes by progressing from simple to complex SED model construction \cite{WALCH2011,SEDFIT2012,PACI2023}. We implement both depth-first search (DFS) and breadth-first search (BFS) strategies. After each node expansion, \texttt{mephisto} evaluates the children based on cumulative residuals, selecting the most promising unexpanded child for subsequent iterations. The SED model refinement and resulting fit quality changes are recorded in \texttt{mephisto}'s temporal memory to avoid redundant efforts. The detail description for the reasoning algorithm can be found in Appendix~\ref{appendix:reasoning}. We note that this approach is fundamentally different from from searching only the best parameters in a fixed model \cite{DIAZ2020,SUN2023b,BRANT2023,CRANMER2020}, as the goal of \texttt{mephisto} here is to come up with various model variants through reasoning instead of just optimizing over all parameters.

\paragraph{Learning Process}\label{subsec:learning} While large language models possess general knowledge of spectral energy distribution, they lack detailed insights into how specific \texttt{CIGALE} modules or parameters affect fit quality. To address this, we've integrated an external knowledge base into \texttt{mephisto}, following approaches in \cite{VOYAGER2023,ZHAO2023,QIAN2024}. This knowledge base is refined through an iterative learning process and can incorporate "intuition" from experienced astronomers. \texttt{mephisto}'s learning process involves a knowledge distillation agent extracting insights from complete fitting histories, a knowledge validation agent assessing each insight's efficacy through experimentation, and a large language model integrating verified knowledge into the base. This validated knowledge is then used as context for future fits, ensuring continuous refinement and more direct action during tree searches. This process effectively allows \texttt{mephisto} to learn from previous SED fitting experiences, improving its performance over time.

\section{Result}\label{sec:result}

We showcase the capabilities of \texttt{mephisto} using the JWST JADES DR2 photometry catalog\cite{JADES2023}\footnote{\url{https://jades-survey.github.io}} in two distinct scenarios. For this analysis, we primarily utilize the GPT-4o model\footnote{\url{https://openai.com/index/hello-gpt-4o/}} as our LLM backbone, as it was one of the top-performing models available when we initiated this study. Our codebase is designed for easy integration of various LLM models. Consequently, we've assessed \texttt{mephisto}'s performance across different LLMs, including the leading open-source model LLaMA-3.1-405B\footnote{\url{https://huggingface.co/meta-llama/Meta-Llama-3.1-405B-Instruct}} and the proprietary GLM-4-0520\footnote{\url{https://open.bigmodel.cn/dev/api/normal-model/glm-4}} (see Appendix~\ref{appendix:llms} for details).

\begin{figure}
\centering
\subfigure[\label{fig:chi2}]{\includegraphics[width=0.45\linewidth]{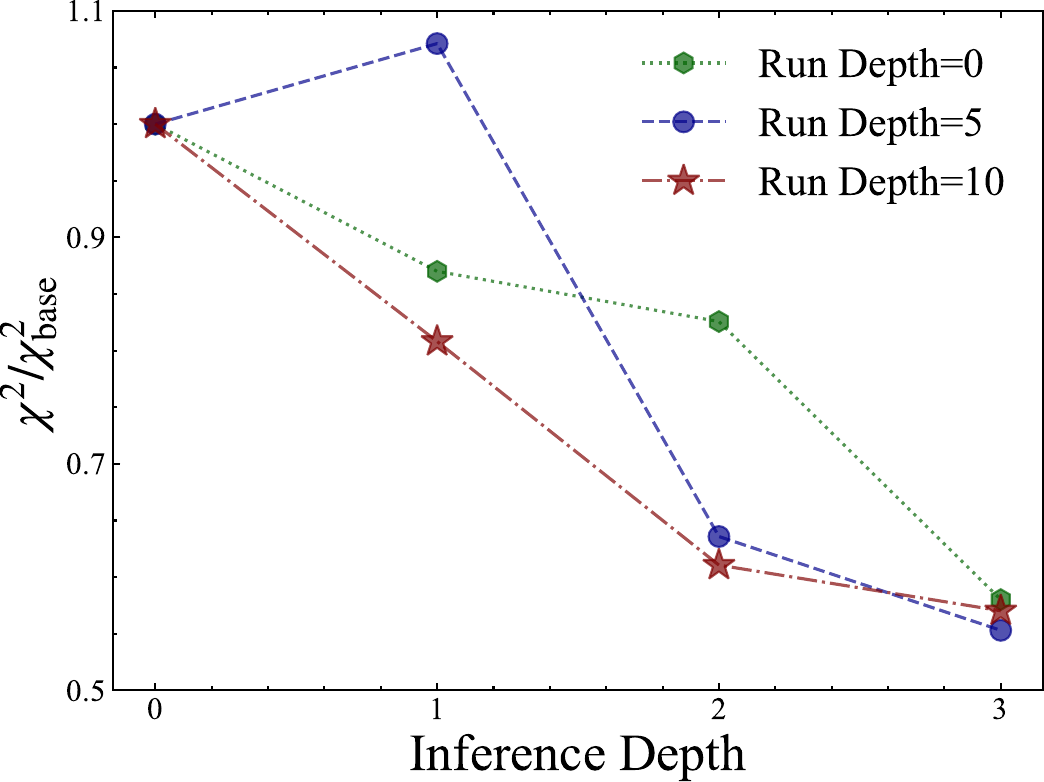}}
\subfigure[\label{fig:result}]{\includegraphics[width=0.45\linewidth]{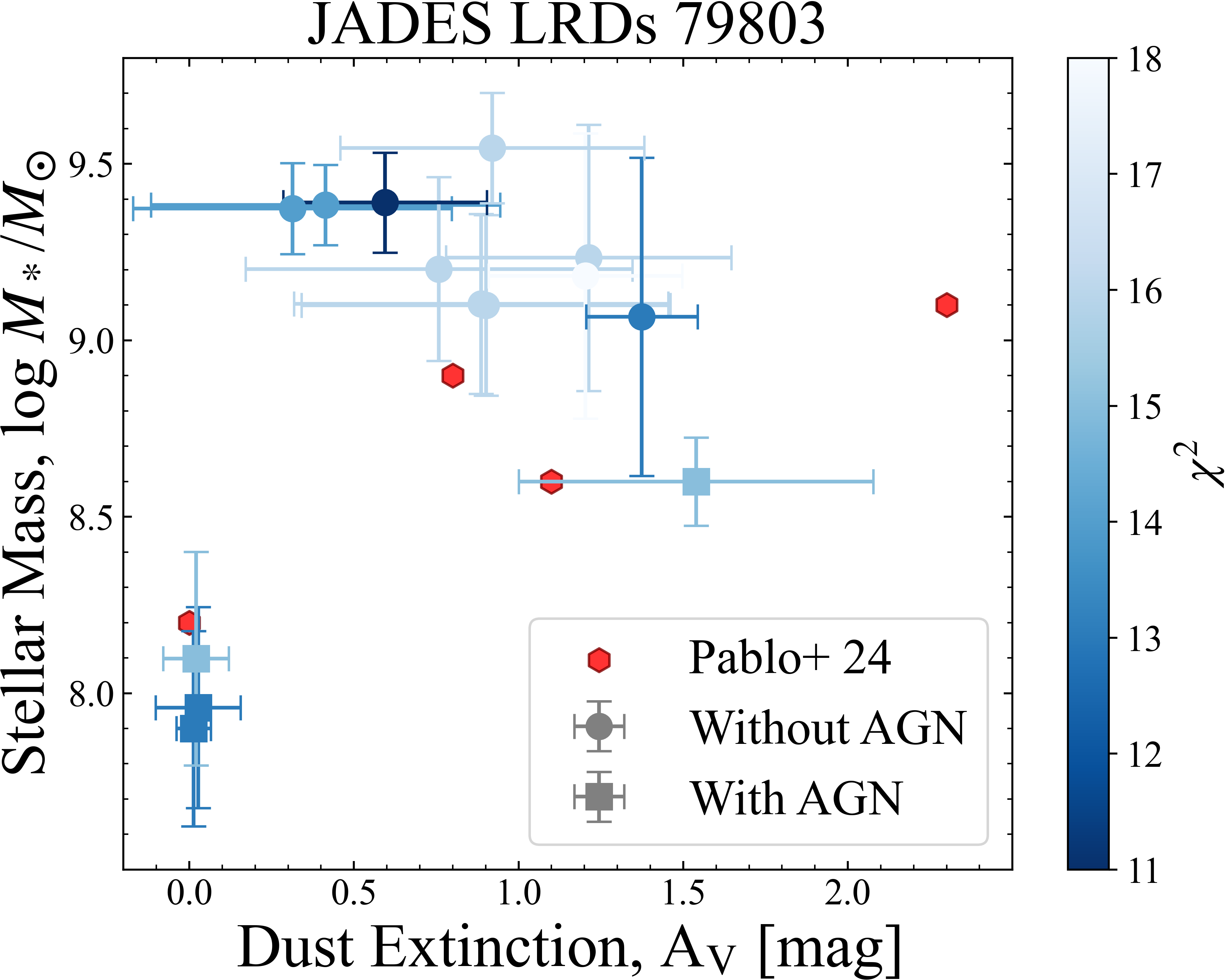}}
\caption{Performance and reasoning capabilities of \texttt{mephisto} on JWST data.  Left: Median $\chi^2$ vs. inference depth for 32 SOM-selected JWST sources. Different lines represent various run depths, demonstrating how \texttt{mephisto}'s accumulated experience through self-play (run depth) improves efficiency (inference depth) in subsequent runs when encountering similar spectra. Right: \texttt{mephisto}'s proposed solutions for a "Little Red Dot" (ID 79803) from the JADES catalog. Blue symbols represent different proposals with CIGALE statistical uncertainties, color-coded by $\chi^2$. Two main scenarios are identified: a dusty star-forming galaxy without AGN and a dust-free galaxy with AGN, consistent with recent literature findings. The derived physical properties are consistent with those proposed in Pablo et al. 2024 \cite{PABLO2024}, shown in red.}
\end{figure}

To show that the model's iterative learning is performing as expected, we first project the JADES data using Self-Organizing Maps (SOM), a dimension reduction technique common in SED analysis \cite{CARR2014,SPEAGLE2019,WRIGHT2020}. We select 32 samples uniformly from the SOM latent space to ensure data diversity. Figure~\ref{fig:chi2} illustrates how the average $\chi^2$ of the fit improves both in terms of inference depth (at a given search) and run depth. Run depth is defined as the number of previous fits performed by \texttt{mephisto}. The results show that the dynamically updated knowledge base continues to enhance search efficiency. This allows \texttt{mephisto} to achieve similar fitting quality with much lower inference depth in subsequent runs when encountering similar spectra.

Beyond fitting the typical JWST data well, we also demonstrate \texttt{mephisto}'s ability to reason about "unknown unknown sources". We tested \texttt{mephisto} on 31 "Little Red Dots" documented in \cite{PABLO2024}. These recently discovered objects were initially hypothesized to be early, dusty galaxies (hence appearing red), but alternative interpretations suggest they could be early galaxies with active galactic nuclei (AGN) \cite{LABBE2023,MATT2024,GENTILE2024,BAGGEN2024}. The right panel of Figure~\ref{fig:chi2} shows the proposed solutions of \texttt{mephisto} for a representative source (with other 30 sources showing similar behavior, see also Appendix~\ref{appendix:lrd}). Error bars indicate the statistical uncertainty from \texttt{CIGALE} for each particular model.

\texttt{mephisto} proposed two primary scenarios: a dusty star-forming galaxy without an AGN, and a dust-free galaxy harboring an AGN (with lower dust extinction $\mathrm{A}_v$) -- the model fits are presented in Appendix~\ref{appendix:lrd}. The derived physical properties are consistent with those reported in \cite{PABLO2024}. Notably, these objects were discovered after GPT-4o's training cutoff date. \texttt{mephisto}'s ability to reason about the nature and debated characteristics of these sources demonstrates its capacity to autonomously conduct "unknown unknown" searches. Our findings show that \texttt{mephisto} can automatically refine SED models to identify diverse explanations for observations without direct human input.

\section{Broader Impact}

This study presents, to our knowledge, the first end-to-end agentic research in astronomy. The SED fitting task serves as an ideal testing ground for LLM agents -- multiple hypotheses often satisfy the limited wavelength information from SED observations, leading to a vast array of potential physical models that preclude exhaustive search. \texttt{mephisto} demonstrates the ability to inherit understanding, distill knowledge, and develop intuition about solver routines through interactions with real data, discovering efficient paths toward multi-step reasoning akin to human researchers.

While SED fitting is a relatively straightforward task, the majority of observed sources from myriad surveys are rarely subjected to detailed human reasoning. Most are triaged based on simple $\chi^2$ criteria through baseline models, potentially causing sources that cannot be explained by current physics to evade detection. The billions of SED sources collected by ongoing and upcoming surveys demand a more streamlined approach. \texttt{mephisto} represents a critical step towards the possibility of exhaustively analyzing all observed sources, potentially uncovering unknown phenomena or competing explanations for observations like the Little Red Dots.

Current limitations to scaling up to the entire sky stem from computational costs. \texttt{mephisto}'s reasoning consumes about 100K tokens, or approximately 1 US dollar per source, making it infeasible for billions of sources. However, as LLMs improve their reasoning ability and costs rapidly decrease, this study marks a crucial step toward fully utilizing all collected astronomical data.

\appendix

\section{\texttt{mephisto}'s Input State}\label{appendix:input}

This appendix provides a detailed example of the input state used by \texttt{mephisto} in its SED fitting process. The input state, formatted in JSON, includes three key components: observational data, the \texttt{CIGALE} model configuration, and the fit quality metrics.

The dataset $d$ encompasses redshift data from JADES and photometric details such as effective wavelengths, bandwidths, fluxes, and signal-to-noise ratios. Due to the current AI limitations in interpreting scientific visuals quantitatively \cite{LILEI2024,YUE2023}, we input data numerically as tuples—a deviation from human researchers' reliance on visual analysis, yet a practical solution for AI agents.

The SED model $m$, constructed in \texttt{CIGALE}, integrates multiple physical components like star formation histories and dust attenuation. The fitting outcome $r$ provides parameter estimates, residuals, computational cost, and fit quality per photometric band, assessed by the agent as good, overestimate, or underestimate. This qualitative method reduces AI numerical hallucinations \cite{YUAN2023,AHN2024}, guiding \texttt{mephisto} with textual cues for each filter evaluation. It also enables \texttt{mephisto} to apply domain knowledge to derive more physical changes, such as modifying dust attenuation physical models for suboptimal optical and UV fits.

The model input is then parsed and run by another agent designed to execute \texttt{CIGALE} codes. The results are evaluated by the reasoning prompt, where they are appended as the new state in the subsequent iteration of the tree search.

Below is an example of the input state structure:

\begin{questionbox}[boxcolor2]
\begin{verbatim}
{"data": {
            "redshift": 0.73,
            "photometry": [
                {
                    "name": "hst.wfc.F435W",
                    "wave_eff": 4337,
                    "bandwidth": 940,
                    "band_location_type": "OPTICAL",
                    "band_width_type": "BROAD",
                    "fit_quality": "good",
                    "signal_to_noise": "reliable"
                },
                ...]},     
        "cigale_model": {
            "sfh": {
                "name": "sfhdelayed",
                "params": {
                "tau_main": [100, 500, 1000, 3000, 5000],
                "age_main": [100, 500, 1000, 3000, 5000],
                "tau_burst": [50],
                "age_burst": [20],
                "f_burst": [0.0]}},...},     
        "cigale_best_parameters": {
            "sfh": {
                "name": "sfhdelayed",
                "params": {
                "tau_main": 100,
                "age_main": 500,
                "tau_burst": 50,
                "age_burst": 20,
                "f_burst": 0.0}},...},     
        "cigale_fit_quality": {
        "num_of_good_fit": 5,
        "sum_chi2": 218,
        "grid_size": 450,
        "cigale_message": "CIGALE run smoothly"}}
\end{verbatim}
\end{questionbox}

\section{\texttt{mephisto}'s Reasoning Process}\label{appendix:reasoning}

This section presents the reasoning prompt used by \texttt{mephisto} to guide its decision-making process in SED fitting. The prompt outlines the rules and knowledge base that \texttt{mephisto} adheres to when modifying CIGALE models to improve fit quality across different photometric bands. The prompt includes guidelines for module selection, parameter grid specification, and mandatory module requirements. It incorporates two key components that enable \texttt{mephisto} to emulate human expert reasoning and learning:

1. \textbf{CIGALE SED Knowledge Base:} This component represents accumulated expertise from previous interactions with data. It enhances \texttt{mephisto}'s ability to improve fit quality for different photometric bands in the CIGALE model. The prompt instructs \texttt{mephisto} to consider this information when designing the CIGALE model. The agent dynamically filled with relevant expert knowledge about SED fitting, allowing \texttt{mephisto} to make informed decisions based on established astrophysical principles and learned insights from past analyses. 

2. \textbf{Temporary Memory:} This feature captures information from the current tree search iteration. It allows \texttt{mephisto} to adapt its strategies within the ongoing analysis. The prompt instructs \texttt{mephisto} to consider these temporary memories to ensure robust diversity and superior fit quality in the updated CIGALE model. The model is updated with information from the current search process, enabling \texttt{mephisto} to refine its approach based on previous suggestions within the same analysis, avoiding redundant proposals.

The integration of the CIGALE SED Knowledge Base and Temporary Memory allows \texttt{mephisto} to generate diverse and reasonable model modifications, combining established principles with both long-term learned insights and short-term adaptive strategies. Importantly, the results of each analysis are not discarded but are instead processed to update the knowledge base. This cycle of learning and integration allows \texttt{mephisto} to continuously refine its knowledge base, improving its performance over time and across different analyses.

The prompt's output format guides \texttt{mephisto} to produce structured, diverse, and reasonable CIGALE model modifications. Each proposed modification includes a "thinking" step, module name, specific choice, and parameter grid values.

\begin{questionbox}[boxcolor1]

$<$\texttt{CIGALE} Documentation$>$

$<$Introduction to User Input$>$

Your task is to analyze user input and adhere to below rules and knowledges to modify the provided \texttt{CIGALE} models to improve the fit quality of different bands.
You should identify the \texttt{CIGALE} module which should affect the fitting quality the most and carefully modify this module, e.g., use another choice, adjust the parameter grid, and etc.

\begin{itemize}
\item Choice Selection: For each module in the model, select either one choice or none. This decision should be based on the need to optimize the model’s fit for the observational data.
\item Parameter Grid Specification: For the selected choice in each module, define a parameter grid that the model will use to fit the data.
\begin{itemize}
\item For discrete parameters, select grid values from a pre-defined list.
\item For continuous parameters, derive grid values within a specified range to ensure a comprehensive exploration of parameter space.
\end{itemize}

\item Mandatory Modules: The model configuration must include specific settings for the following modules:
\begin{itemize}
\item sfh (Star Formation History): This module is essential for modeling the rate at which a galaxy forms stars over time.
\item ssp (Simple Stellar Population): This module is crucial for understanding the collective properties of stars in a galaxy that formed at the same time and with the same metallicity.
\end{itemize}
\item For 'imf' parameter in 'bc03' and 'm2005', it should be 0 or 1, not a list, i.e., "imf": 1 are accepteable, "imf": [1] are foridden
\item  For 'disk\_type' parameter in 'fritz2006' and 'skirtor2016' it should be 0 or 1, not a list, i.e., "disk\_type": 0 are acceptable, "disk\_type": [1] are forbidden
\end{itemize}

\#\#CIGALE SED Knowledge\#\#

To enhance the fit quality for different photometric bands in the CIGALE model, consider the following addtional information when designing your CIGALE model.

\%\%KNOWLEDGE\%\%

$\quad$

\#\#Temporary Memory\#\#

Please take into account the following temporary memories to ensure a robust diversity and superior fit quality in the updated CIGALE model:

\%\%MEMORY\%\%

$\quad$

\#\#Expected Output Format\#\#

Your output format should be structured as below list constructed by four diverse and reasonable CIGALE model modifications:

\begin{verbatim}
[
    {
        "thinking": "thinking for CIGALE model modification 0 here"
        "module": "module name for CIGALE model modification 0 here",
        "name": "module choice name for CIGALE model modification 0 here",
        "parameters": [
            "parameter 1": ["parameter 1 grid here"],
            "parameter 2": ["parameter 2 grid here"],
            ...
            "parameter n": ["parameter n grid here"]
        ]
    },
    ...
]
\end{verbatim}
\end{questionbox}

To better understand the reasoning process, we present a detailed pseudocode of \texttt{mephisto}'s adaptive search strategy for exploring possible model modifications. 

At its core, the algorithm iterates between two search strategies: depth-first search (DFS) and breadth-first search (BFS). This hybrid approach allows for a more comprehensive and balanced exploration of the solution space. The algorithm begins with an initial state and alternates between DFS and BFS to generate and evaluate new states until a termination condition is met. This alternation helps to overcome the limitations of each individual strategy:

1. Depth-First Search (DFS): Allows for quick exploration of deep branches in the solution space, potentially finding highly specialized configurations. It uses a stack data structure, which prioritizes exploring the most recently added states first.

2. Breadth-First Search (BFS): Ensures a more balanced exploration across all branches, preventing the algorithm from overlooking potentially valuable shallow configurations. It uses a queue data structure, which prioritizes exploring states in the order they were added.

By iterating between these two strategies, the algorithm can adapt to the specific characteristics of the problem at hand, balancing between deep exploration and broad coverage of the solution space.

Throughout this process, the algorithm dynamically incorporates both existing knowledge and accumulated memory to inform its decisions, continuously updating these components as it progresses. This adaptive approach allows the algorithm to learn and improve its performance over time.

In the following, we have:

\begin{itemize}
    \item $N_{\text{max}}$: Maximum number of total states to explore
    \item $D_{\text{max}}$: Maximum depth of the search tree
    \item $s_0$: Initial state (root of the search tree)
    \item $\mathcal{K}$: Knowledge base
    \item $\mathcal{M}$: Memory of explored states
    \item $R$: Reasoning agent that generates new states
    \item $E$: Evaluation function that assesses new states
\end{itemize}

\begin{algorithm}
\begin{algorithmic}[1]
\STATE \textbf{Input:} $s_0$, $\mathcal{K}$, $\mathcal{M}=\{\}$, $R$, $E$, $D_{\text{max}}$, $N_{\text{max}}$
\STATE Initialize stack $S = [s_0]$   \#$\,$Stack for DFS
\STATE $\text{total\_states} = 1$  \#$\,$Count of explored states
\WHILE{$S \neq [\,]$ AND $\text{total\_states} < N_{\text{max}}$}
    \STATE $s = S.\text{pop}()$  Get the most recently added state
    \STATE $s'_1,\ldots,s'_n = R(s, \mathcal{K}, \mathcal{M})$  $\,$\#$\,$Generate new states
    \STATE $s''_1,\ldots,s''_n = E(s'_1,\ldots,s'_n)$  $\,$\#$\,$Evaluate new states
    \FOR{$s''$ in $s''_1,\ldots,s''_n$}
        \STATE $s.\text{add\_child}(s'')$  $\,$\#$\,$Add new state to the tree
        \IF{$s''.\text{depth} < D_{\text{max}}$}
            \STATE $S.\text{push}(s'')$  $\,$\#$\,$Add to stack if not too deep
            \STATE $\text{total\_states} \gets \text{total\_states} + 1$
        \ENDIF
    \ENDFOR
    \STATE $\mathcal{M}.\text{update}(s, s''_1,\ldots,s''_n)$  $\,$\#$\,$Update memory
\ENDWHILE
\STATE $\mathcal{K} \gets \mathcal{K}.\text{update}(\mathcal{M})$  $\,$\#$\,$Update knowledge base
\RETURN $\text{summarize}(s_0)$  $\,$\#$\,$Summarize the entire search tree
\end{algorithmic}
\caption{Depth-First Search Reasoning Process}
\end{algorithm}

\begin{algorithm}
\begin{algorithmic}[1]
\STATE \textbf{Input:} $s_0$, $\mathcal{K}$, $\mathcal{M}=\{\}$, $R$, $E$, $D_{\text{max}}$, $N_{\text{max}}$
\STATE Initialize queue $Q = [s_0]$  $\,$\#$\,$Queue for BFS
\STATE $\text{total\_states} = 1$  $\,$\#$\,$ Count of explored states
\WHILE{$Q \neq [\,]$ AND $\text{total\_states} < N_{\text{max}}$}
    \STATE $s = Q.\text{dequeue}()$  $\,$\#$\,$ Get the earliest added state
    \STATE $s'_1,\ldots,s'_n = R(s, \mathcal{K}, \mathcal{M})$  $\,$\#$\,$ Generate new states
    \STATE $s''_1,\ldots,s''_n = E(s'_1,\ldots,s'_n)$  $\,$\#$\,$ Evaluate new states
    \FOR{$s''$ in $s''_1,\ldots,s''_n$}
        \STATE $s.\text{add\_child}(s'')$  $\,$\#$\,$ Add new state to the tree
        \IF{$s''.\text{depth} < D_{\text{max}}$}
            \STATE $Q.\text{enqueue}(s'')$  $\,$\#$\,$ Add to queue if not too deep
            \STATE $\text{total\_states} \gets \text{total\_states} + 1$
        \ENDIF
    \ENDFOR
    \STATE $\mathcal{M}.\text{update}(s, s''_1,\ldots,s''_n)$  $\,$\#$\,$ Update memory
\ENDWHILE
\STATE $\mathcal{K} \gets \mathcal{K}.\text{update}(\mathcal{M})$  $\,$\#$\,$ Update knowledge base
\RETURN $\text{summarize}(s_0)$  $\,$\#$\,$ Summarize the entire search tree
\end{algorithmic}
\caption{Breadth-First Search Reasoning Process}
\end{algorithm}

\section{\texttt{mephisto}'s Learning Process}\label{appendix:learn}

This section details the learning process implemented in \texttt{mephisto}, focusing on how the system acquires, validates, and integrates knowledge to enhance its performance in Spectral Energy Distribution (SED) fitting. The external knowledge base in \texttt{mephisto} is formatted in natural language, as shown in Figure~\ref{fig:schema}. This knowledge is crucial for navigating the complex hypothesis space of SED models, illustrated in Figure~\ref{fig:chi2}.

During the reasoning process, as described in the previous appendix, \texttt{mephisto} generates various suggestions for modifying the current SED model to address discrepancies between predictions and empirical data. These refined models are computed using \texttt{CIGALE} and assessed for fit quality improvement. The interaction history is logged and processed by a learning agent, which extracts insights to optimize future SED model designs.

To ensure the quality and relevance of the acquired knowledge, we have implemented a robust validation and distillation process:

1. Knowledge Extraction: The learning agent analyzes the interaction history to synthesize critical insights for improving CIGALE models. These insights are formulated as universally relevant knowledge that can address inherent design flaws in the models.

2. Knowledge Validation: A validation agent compiles each piece of knowledge into a logical expression. These expressions are then evaluated using an eval\_state function to determine their applicability to different observation states. This step ensures that the knowledge is consistent with the provided background information and can be applied across various scenarios.

3. Impact Assessment: The validated knowledge is evaluated based on its impact on reasoning outcomes across multiple sources. This step helps identify knowledge that consistently improves model performance.

4. Knowledge Filtering: Following the impact assessment, knowledge with zero or negative impact is filtered out, ensuring that only beneficial knowledge is retained.

5. Knowledge Integration: The remaining beneficial knowledge is integrated into the existing knowledge base through an optimization agent. This agent analyzes discrepancies between new and existing knowledge, synthesizing them into a cohesive whole.

This iterative learning process allows \texttt{mephisto} to continuously refine its knowledge base, improving its ability to navigate the complex hypothesis space of SED models and generate more accurate fits over time.

The following prompt illustrates the knowledge extraction component of this learning process:

\begin{questionbox}[boxcolor2]
\#\# Task: Refine Key Insights from Interaction History for CIGALE Model Enhancement

$\quad$

\#\#\# Objective:

$\quad$

Your task is to synthesize critical insights from the supplied interaction history, which will be instrumental in shaping the development of upcoming CIGALE models. The primary focus should be on improving their overall performance and effectiveness.\\

The insights you derive must be universally relevant and capable of addressing any inherent design flaws in the models.\\

Your analysis is expected to be consistent with the details provided in the interaction history.\\

$\quad$
<Examples>
$\quad$

\#\#\# Output Format:

$\quad$

List of distilled Knowledge:
$\quad$
\begin{verbatim}
["knowledge 1", "knowledge 2", ...]
\end{verbatim}
\end{questionbox}

The following prompt illustrates the knowledge validation component of this learning process:

\begin{questionbox}[boxcolor2]
Objective: Derive a logical expression from the given knowledge on spectral energy distribution fitting in galaxy physics to determine the applicability of the knowledge to a specified observation state.\\

Background Information:

<Background Information Here>\\

$\quad$

Evaluation Function: Use the eval\_state function to assess observations:

def eval\_state(state, feature, pattern):

$\quad$

    Parameters:

$\quad$

    - state: Identifier for a multi-band observation.

$\quad$

    - feature: Band name or tuple (Band Location Type, Band Width Type). 'ANY' can be used as a wildcard.

$\quad$

    - pattern: Fit quality criterion (good, overestimated, underestimated).

$\quad$

    Returns:

$\quad$

    - True if the state meets the feature with the specified pattern, otherwise False.

$\quad$

Task: Formulate a logical expression using the eval\_state function to identify multi-band observations where the provided knowledge is applicable.
Note: For intricate knowledge, merge conditions using logical operators such as OR (or) and AND (and).
Note: Employ parentheses to structure complex logical expressions as needed.

$\quad$

<Examples>\\

$\quad$

Output Format:

\begin{verbatim}
{
    "thinking": "detail your thinking here",
    "expression": "expression here"
}
\end{verbatim}
\end{questionbox}

The following prompt illustrates the knowledge integration component of this learning process:

\begin{questionbox}[boxcolor2]
Your task is to maintain a knowledge base used to guide the design of CIGALE SED models. \\

You will receive a knowledge base and a set of new knowledge. \\

You need to distill the new information and integrate it into the existing knowledge base.\\

\#\# Guidelines:

1. If the newly added knowledge does not conflict with any existing knowledge, add it to the knowledge base.

2. If the newly added knowledge conflicts with an existing piece of knowledge, revise the existing knowledge to make them as consistent as possible.

3. If the newly added knowledge is essentially the same as an existing piece of knowledge, organize and merge them.

4. Make the distilled knowledge consise and informative.\\

\#\# Input Format:

\begin{verbatim}
{
    "BaseKnowledge": [
        "a list of base knowledge here"
    ],
    "KnowledgeGradient": [
        "a list of new knowledge here"
    ]
}
\end{verbatim}

\#\# Output Format:

\begin{verbatim}
{
    "thinking": "Describing the difference 
    between BaseKnowledge and KnowledgeGradient first, 
    and analyze how to combine the two sets of knowledge.",
    "UpdatedKnowledgeBase": [
        "a list of updated knowledge "
    ]
}
\end{verbatim}
\end{questionbox}

\section{Using Different LLM Backbones for \texttt{mephisto}}\label{appendix:llms}

While we have demonstrated that using GPT-4o is sufficient to achieve successful self-play reinforcement learning in our open-world SED fitting setting, the current pipeline costs approximately one dollar per source for all the detailed reasoning and checking, as shown in the previous appendices. A key question, therefore, is whether there might be more affordable options that could perform the same job.

Although a comprehensive comparison of all LLMs is beyond the scope of this paper, we conducted a preliminary study using 32 JADES galaxies, selected via SOM as detailed in Section~\ref{sec:result}, while varying the underlying LLM. The evaluation encompassed GPT-4o (used in the main text), LLaMA-3.1-405B (a state-of-the-art open-source LLM), and GLM-4-0520.

To ensure a fair comparison, each agent was provided with comprehensive knowledge distilled from the interaction history by GPT-4o, minimizing the impact of differences in internal domain knowledge among the LLMs. The SED fitting task requires the LLM to accurately interpret \texttt{CIGALE}'s documentation, adhere to prompt instructions, comprehend input data, and effectively utilize provided knowledge and memory—all of which demand advanced long-context reasoning capabilities. We note that this remains a fair comparison because such knowledge distillation, in principle, only needs to be conducted once with the best-performing model, and the knowledge, in the form of natural language, can be shared with more affordable models without requiring knowledge base updates.

\begin{figure}\label{fig:llms}
\centering
\subfigure[Median $\chi^2$ vs. Inference Depths\label{fig:median}]{\includegraphics[width=0.45\linewidth]{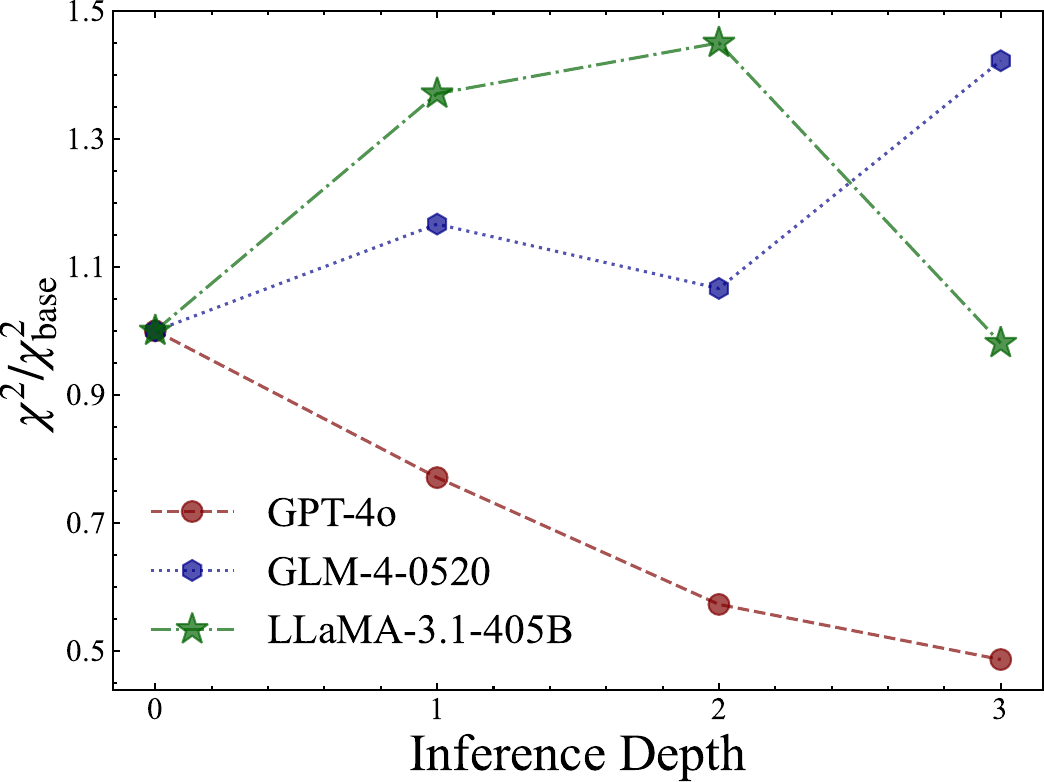}}
\subfigure[Minimal $\chi^2$ vs. Inference Depths\label{fig:minimal}]{\includegraphics[width=0.45\linewidth]{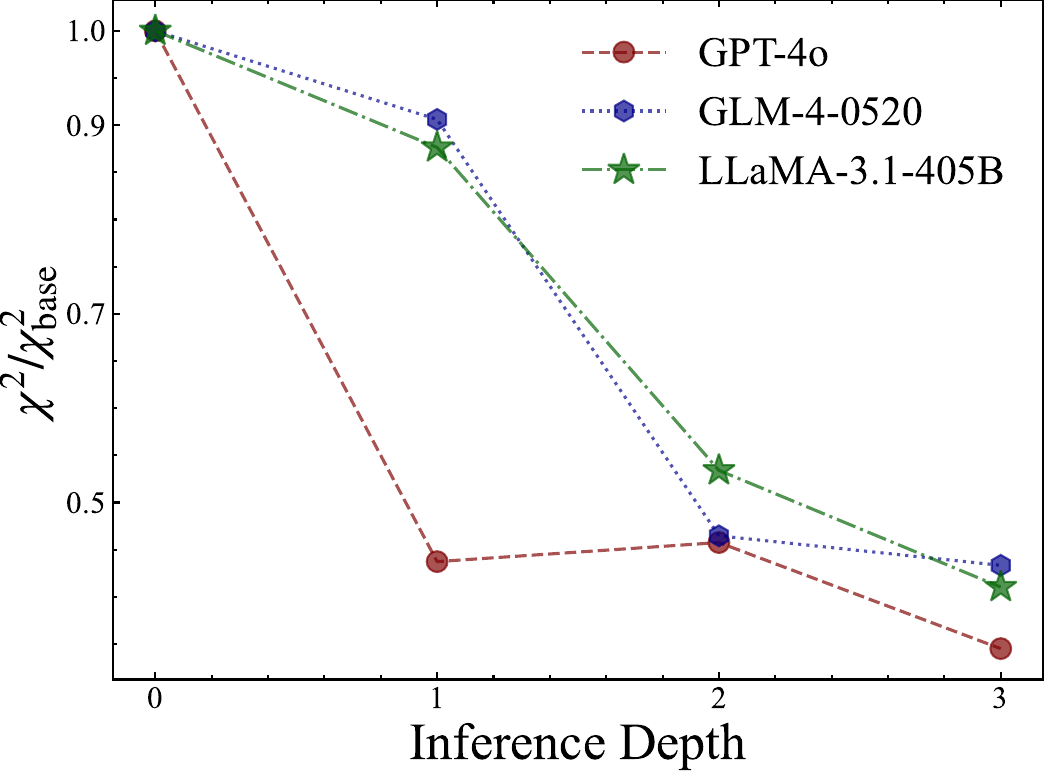}}
\caption{Reasoning capabilities of \texttt{mephisto} with different LLMs on JWST data. Left: Median $\chi^2$ from all proposals vs. inference depth, averaged over 32 SOM-selected JWST sources. $\chi^2_{base}$ denotes $\chi^2$ of the root state. GPT-4o shows the best average performance, effectively utilizing the extracted knowledge base to improve proposals consistently. LLaMA-3.1-405B and GLM-4-0520 struggle to leverage the knowledge base, resulting in poorer average performance. Right: Best $\chi^2$ among all the proposals vs. inference depth, averaged over 32 SOM-selected JWST sources, showing the best proposal for each source. While GPT-4o demonstrates consistent improvement, Llama-3.1-405B and GLM-4-0520 occasionally find good solutions for some galaxies, despite their overall inefficiency in using the knowledge base. This highlights the importance of cost-effectiveness in practical applications and the need for detailed benchmarking in astronomical research scenarios.}
\end{figure}

Interestingly, as illustrated in the left panel of Figure~\ref{fig:median}, which shows the median $\chi^2$ over all the proposals, averaged over all 32 galaxies, and the right panel, which shows the best $\chi^2$ among all the proposals, averaged over the 32 galaxies, only GPT-4o is able to harness the extracted knowledge and lead to better inference at different inference depths, benefiting from the knowledge on average. While LLaMA-3.1-405B and GLM-4-0520 have shown the ability to find the solution among some of their proposals as the inference depth increases (as seen in the best $\chi^2$), on average, they fail to effectively utilize the extracted knowledge base. This result suggests that despite converging performance on standard natural language processing benchmarks like MMLU \cite{MMLU2020}, LLM capabilities can still vary considerably in open-world, long-context reasoning scenarios. However, we note that it remains possible that even in the form of natural language, there might be subtle differences in the preferences of different LLMs in terms of the format and style of the knowledge, which could lead to the GPT-4o extracted knowledge being more applicable to GPT-4o. This is an area we aim to explore further.

Nonetheless, taken at face value, our findings contrast with those of Ting et al. 2024 \cite{TING2024}, who evaluated various LLMs on astronomical question answering. While their study highlighted the internal astronomical knowledge of LLMs, our approach bridges the domain expertise gap through an external knowledge base, focusing more on the reasoning abilities required for practical research challenges. Consequently, LLaMA-3.1-405B, which performed on par with GPT-4o in Ting et al.'s study, shows diminished efficacy in our task. Interestingly, GLM-4-0520, which underperformed compared to LLaMA-3.1-405B in Ting et al.'s research, demonstrates comparable performance to LLaMA-3.1-405B in our experimental framework. This further demonstrates the importance of detailed benchmarking for different use cases.

\texttt{mephisto} may thus serve as a valuable benchmark for assessing the reasoning abilities of various LLMs in astronomical research scenarios, potentially being the first of its kind in this domain.

\section{SED Fitting Result on Little Red Dots}\label{appendix:lrd}

This appendix presents the SED fitting results obtained by \texttt{mephisto} for five representative JADES Little Red Dot (LRD) galaxies, demonstrating its ability to reason about complex, multimodal solutions in a manner akin to human researchers.

Little Red Dots, recently discovered by the James Webb Space Telescope, have sparked debate in the astronomical community. Initially thought to be early, dusty galaxies, they may alternatively be early galaxies with active galactic nuclei (AGN) \cite{PABLO2024,MATT2024,GENTILE2024,BAGGEN2024}. This uncertainty underscores the challenges in interpreting high-redshift galaxy observations and the need for sophisticated, human-like analysis techniques.

For each LRD, \texttt{mephisto} has identified two distinct, physically plausible solutions:
\begin{enumerate}
    \item A dusty star-forming galaxy without an AGN
    \item A relatively dust-free galaxy hosting an AGN
\end{enumerate}

Crucially, \texttt{mephisto} arrived at these solutions through a process of reflection and model changes, not simply via hyperparameter search. This approach demonstrates its capacity for "unknown unknown" searches, proposing and evaluating multiple hypotheses to explain challenging observations.

The figures illustrate these distinct solutions, showcasing \texttt{mephisto}'s ability to navigate the complex hypothesis space of SED modeling. Each subfigure provides key derived parameters: dust attenuation ($A_v$), stellar mass ($M_*$), and AGN fraction, highlighting the fundamental differences between the proposed scenarios.

By identifying multiple valid solutions, \texttt{mephisto} contributes meaningfully to ongoing astronomical debates, providing a comprehensive view of possible physical scenarios for these objects. The contrasting solutions highlight the degeneracies inherent in SED fitting for high-redshift objects with limited photometric data, mimicking the careful consideration human experts apply to such challenging cases.

The range of physical parameters explored - from nearly dust-free to heavily obscured galaxies, and from pure star-forming systems to AGN-dominated ones - demonstrates \texttt{mephisto}'s flexibility in hypothesis generation. This capability is key to performing "unknown unknown" searches, potentially uncovering unexpected physical scenarios that human researchers might overlook.

These results showcase \texttt{mephisto}'s potential as a powerful tool for astronomical research, capable of not only fitting SEDs but also engaging in exploratory data analysis and hypothesis generation traditionally performed by human experts.

\begin{figure}[htbp!]
    \centering
    \subfigure[$A_v=0.03\pm0.13\,\mathrm{mag}$ ; $\log M_*/M_{\odot}=7.96\pm0.28$ ; $\mathrm{frac}_{\mathrm{AGN}}=0.99$ ; $z_{spec}=5.4007$]{\includegraphics[width=0.68\linewidth]{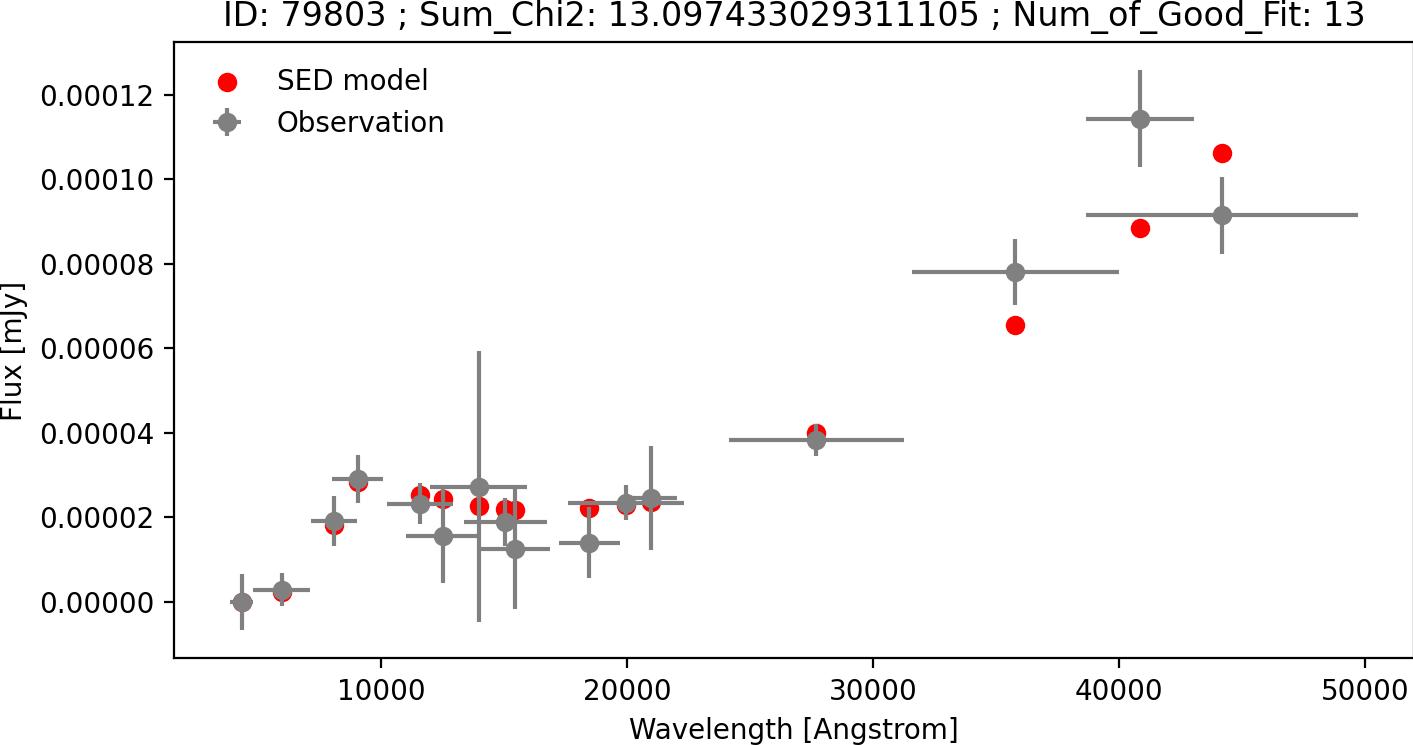}}
    \subfigure[$A_v=0.92\pm0.46\,\mathrm{mag}$ ; $\log M_*/M_{\odot}=9.54\pm0.15$ ; $\mathrm{frac}_{\mathrm{AGN}}=0$ ; $z_{spec}=5.4007$]{\includegraphics[width=0.68\linewidth]{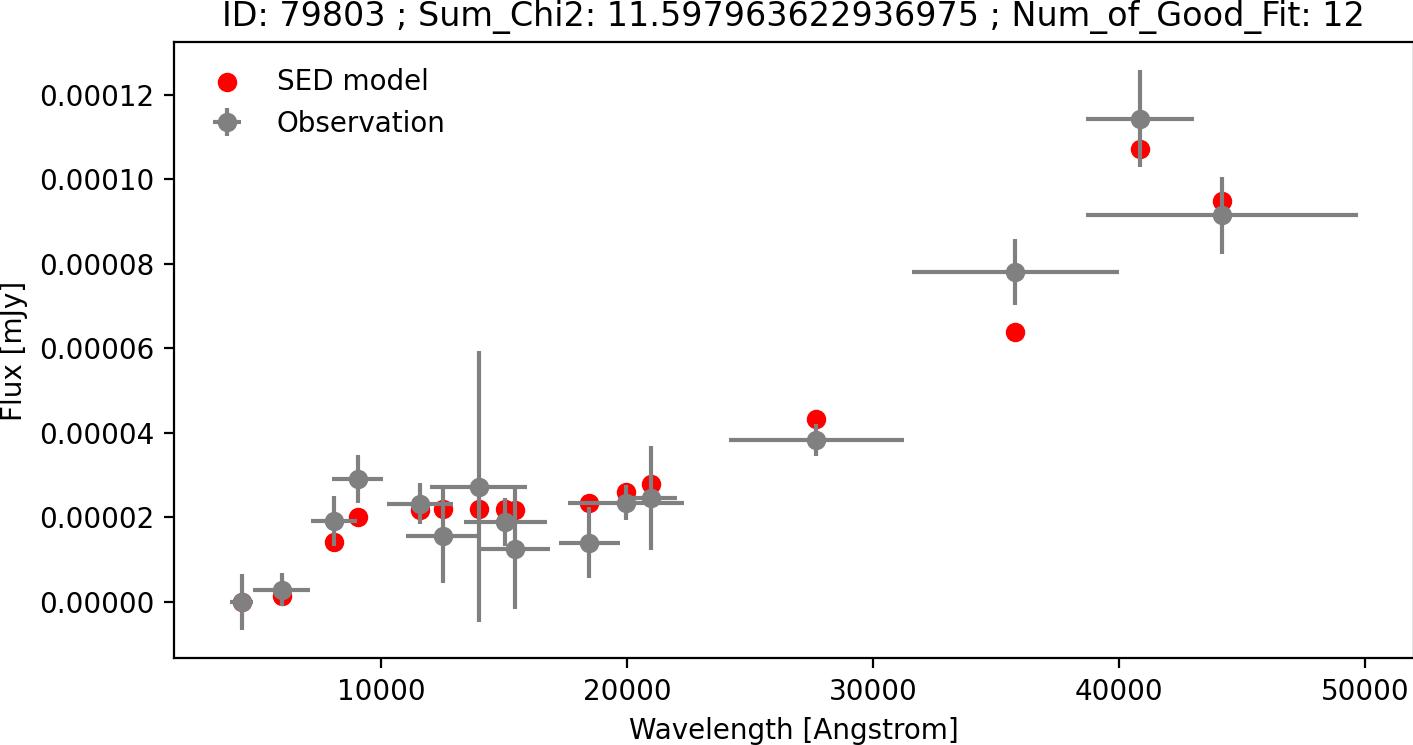}}
    \caption{SED Fitting Results for JADES LRD 79803. Panel (a) shows a solution favoring a dust-free galaxy with a dominant AGN ($\mathrm{frac}_{\mathrm{AGN}} = 0.99$), while panel (b) presents an alternative scenario of a dustier ($A_v = 0.92$) star-forming galaxy without AGN contribution. These contrasting solutions, both providing good fits to the observational data, exemplify the current debate in the literature regarding the nature of Little Red Dots.}
    \label{fig:lrd_79803}
\end{figure}

\begin{figure}[htbp!]
    \centering
    \subfigure[$A_v=10.1\pm2.2\,\mathrm{mag}$ ; $\log M_*/M_{\odot}=12.7\pm0.12$ ; $\mathrm{frac}_{\mathrm{AGN}}=0.99$ ; $z_{spec}=7.6641$]{\includegraphics[width=0.68\linewidth]{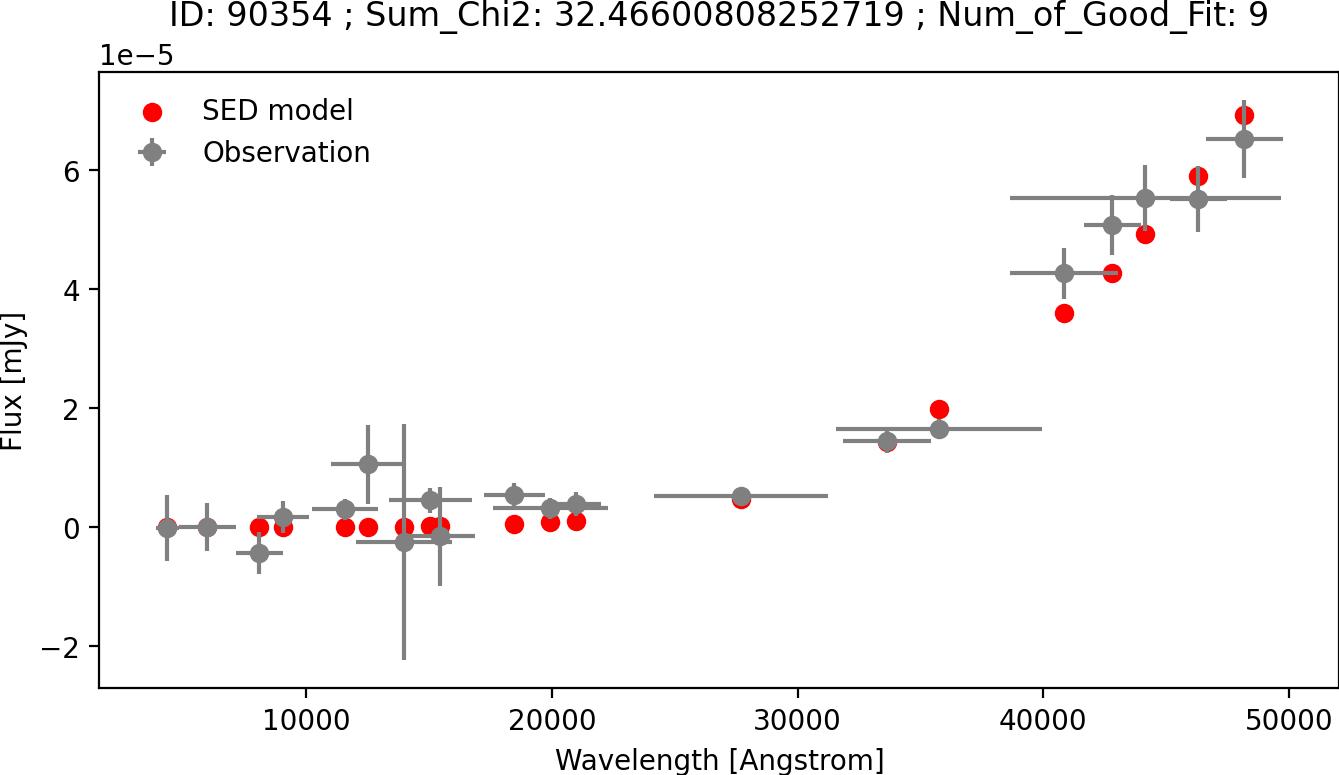}}
    \subfigure[$A_v=3.01\pm0.53\,\mathrm{mag}$ ; $\log M_*/M_{\odot}=10.1\pm0.12$ ; $\mathrm{frac}_{\mathrm{AGN}}=0$ ; $z_{spec}=7.6641$]{\includegraphics[width=0.68\linewidth]{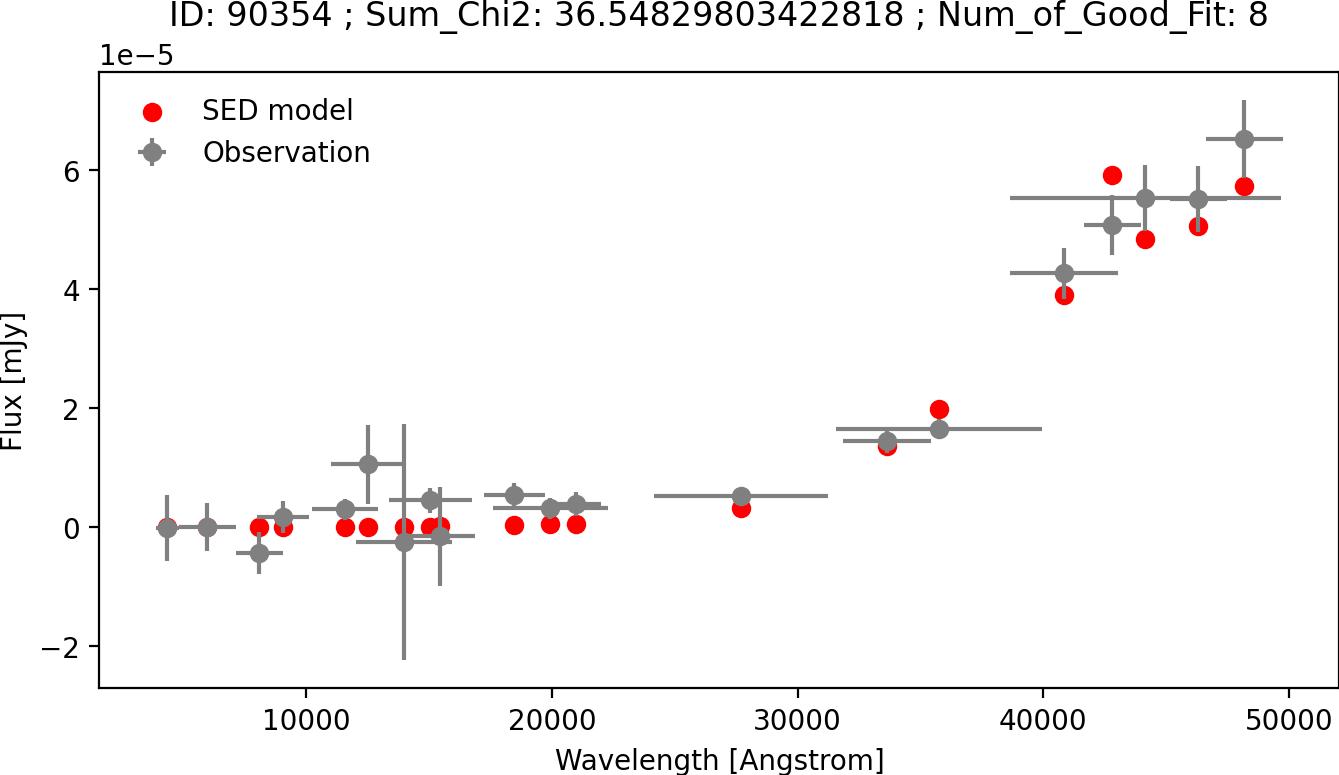}}
    \caption{SED Fitting Results for JADES LRD 90354. Panel (a) depicts a heavily dust-obscured ($A_v = 10.1$) galaxy with a strong AGN component, while panel (b) shows a less dusty ($A_v = 3.01$) star-forming galaxy without AGN.}
    \label{fig:lrd_90354}
\end{figure}

\begin{figure}[htbp!]
    \centering
    \subfigure[$A_v=0.77\pm0.65\,\mathrm{mag}$ ; $\log M_*/M_{\odot}=6.68\pm1.12$ ; $\mathrm{frac}_{\mathrm{AGN}}=0.99$ ; $z_{spec}=5.0047$]{\includegraphics[width=0.68\linewidth]{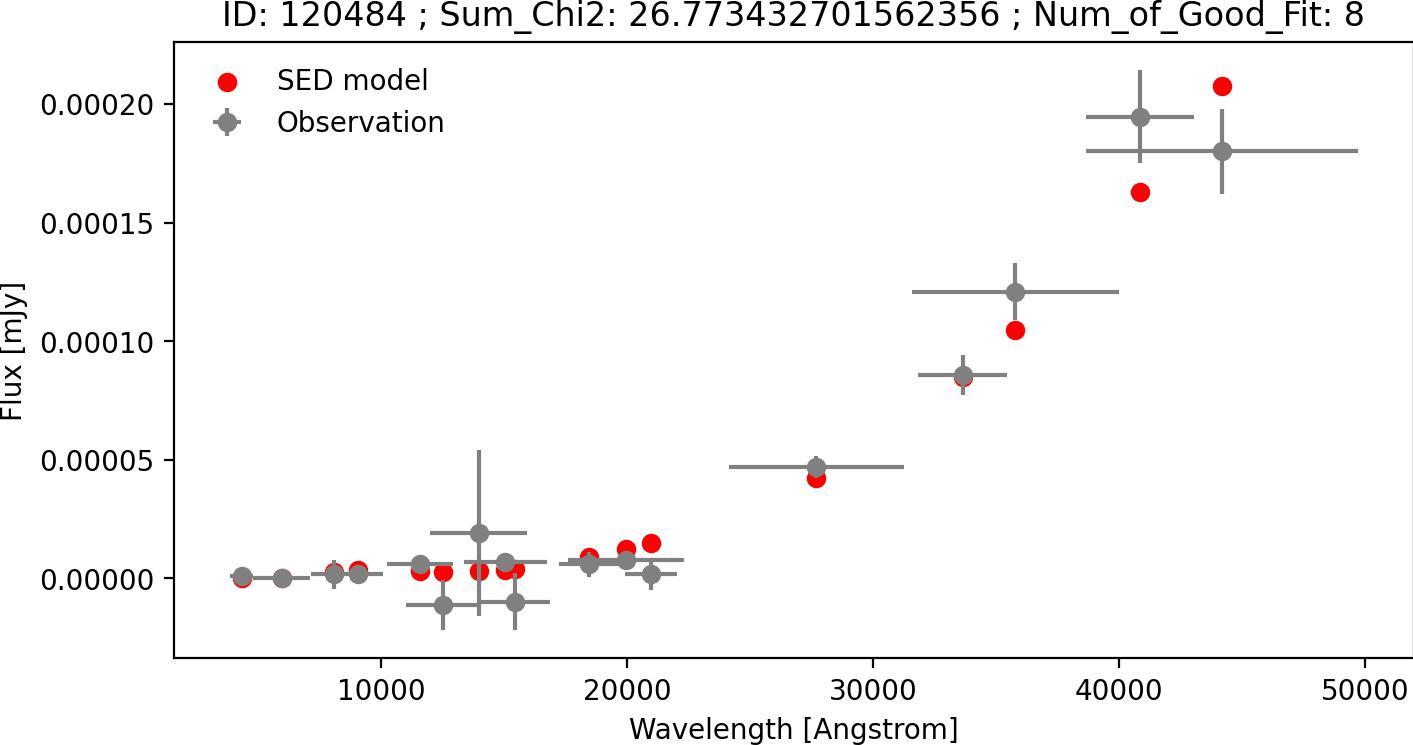}}
    \subfigure[$A_v=0.92\pm0.46\,\mathrm{mag}$ ; $\log M_*/M_{\odot}=9.55\pm0.16$ ; $\mathrm{frac}_{\mathrm{AGN}}=0$ ; $z_{spec}=5.0047$]{\includegraphics[width=0.68\linewidth]{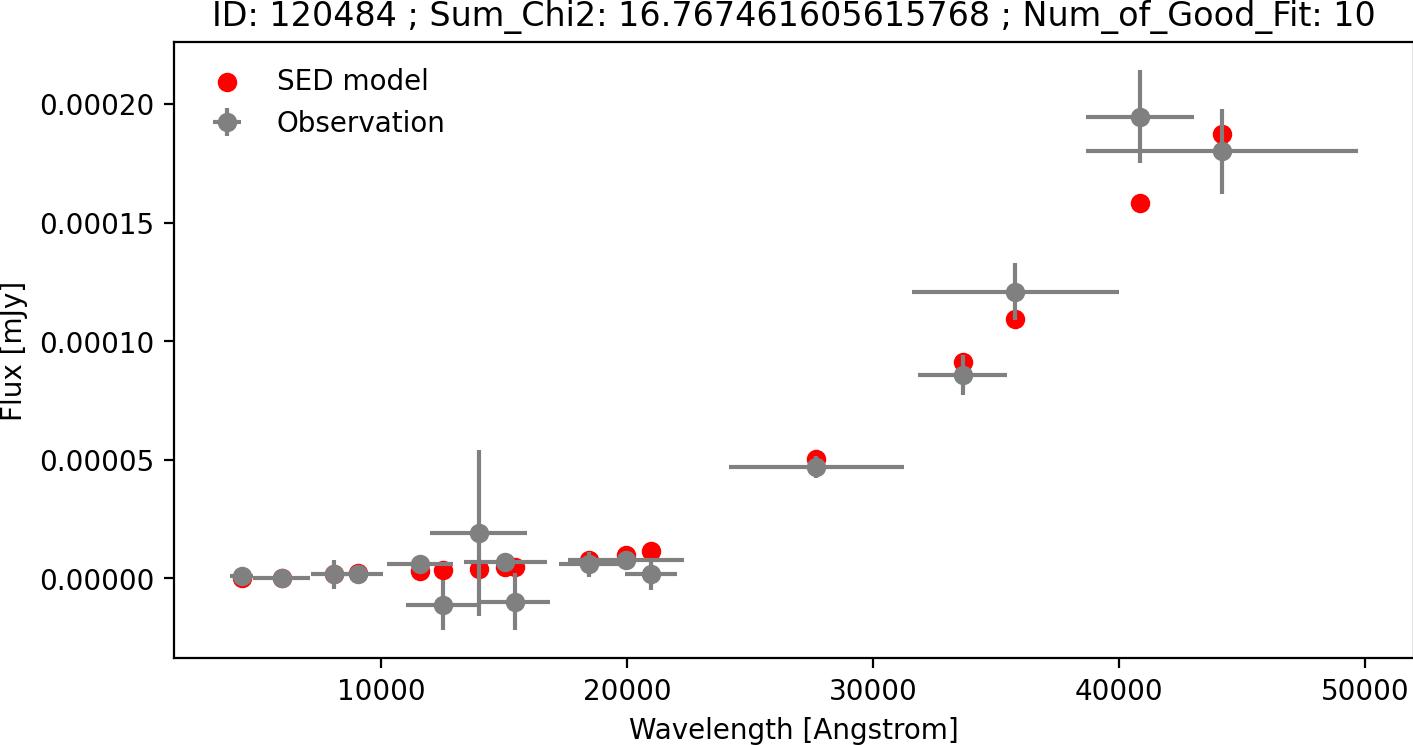}}
    \caption{SED Fitting Results for JADES LRD 120484. Panel (a) presents a low-mass galaxy dominated by AGN emission, while panel (b) shows a more massive, purely star-forming galaxy.}
    \label{fig:lrd_120484}
\end{figure}

\begin{figure}[htbp!]
    \centering
    \subfigure[$A_v=0.07\pm0.65\,\mathrm{mag}$ ; $\log M_*/M_{\odot}=8.70\pm0.15$ ; $\mathrm{frac}_{\mathrm{AGN}}=0.99$ ; $z_{spec}=8.6751$]{\includegraphics[width=0.68\linewidth]{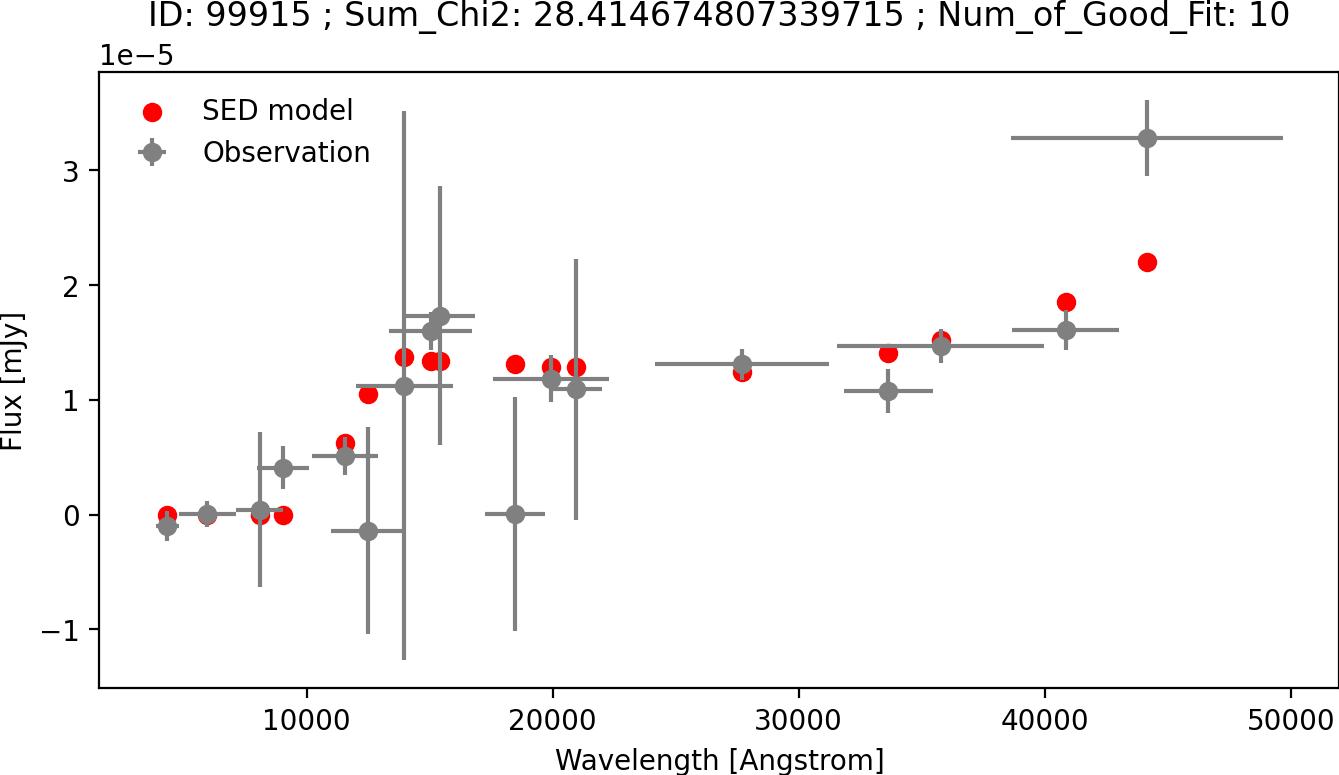}}
    \subfigure[$A_v=0.46\pm0.56\,\mathrm{mag}$ ; $\log M_*/M_{\odot}=8.2\pm0.27$ ; $\mathrm{frac}_{\mathrm{AGN}}=0$ ; $z_{spec}=8.6751$]{\includegraphics[width=0.68\linewidth]{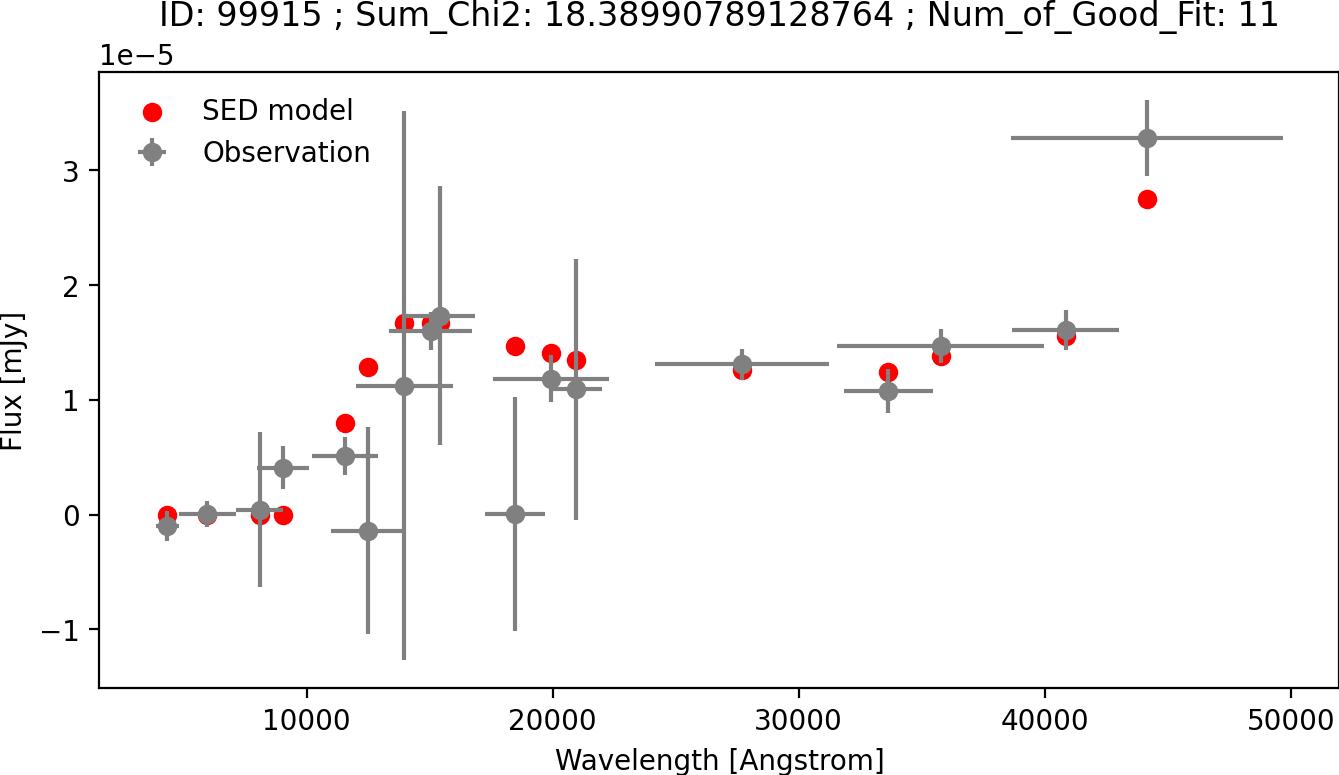}}
    \caption{SED Fitting Results for JADES LRD 99915. Both panels show relatively low-dust solutions, with panel (a) featuring a strong AGN component and panel (b) representing a purely star-forming galaxy. This case demonstrates that even with similar dust attenuation, the presence or absence of an AGN can significantly alter the interpretation of the galaxy's nature.}
    \label{fig:lrd_99915}
\end{figure}

\begin{figure}[htbp!]
    \centering
    \subfigure[$A_v=1.35\pm1.12\,\mathrm{mag}$ ; $\log M_*/M_{\odot}=9.22\pm0.61$ ; $\mathrm{frac}_{\mathrm{AGN}}=0.99$ ; $z_{phot}=5.5$]{\includegraphics[width=0.68\linewidth]{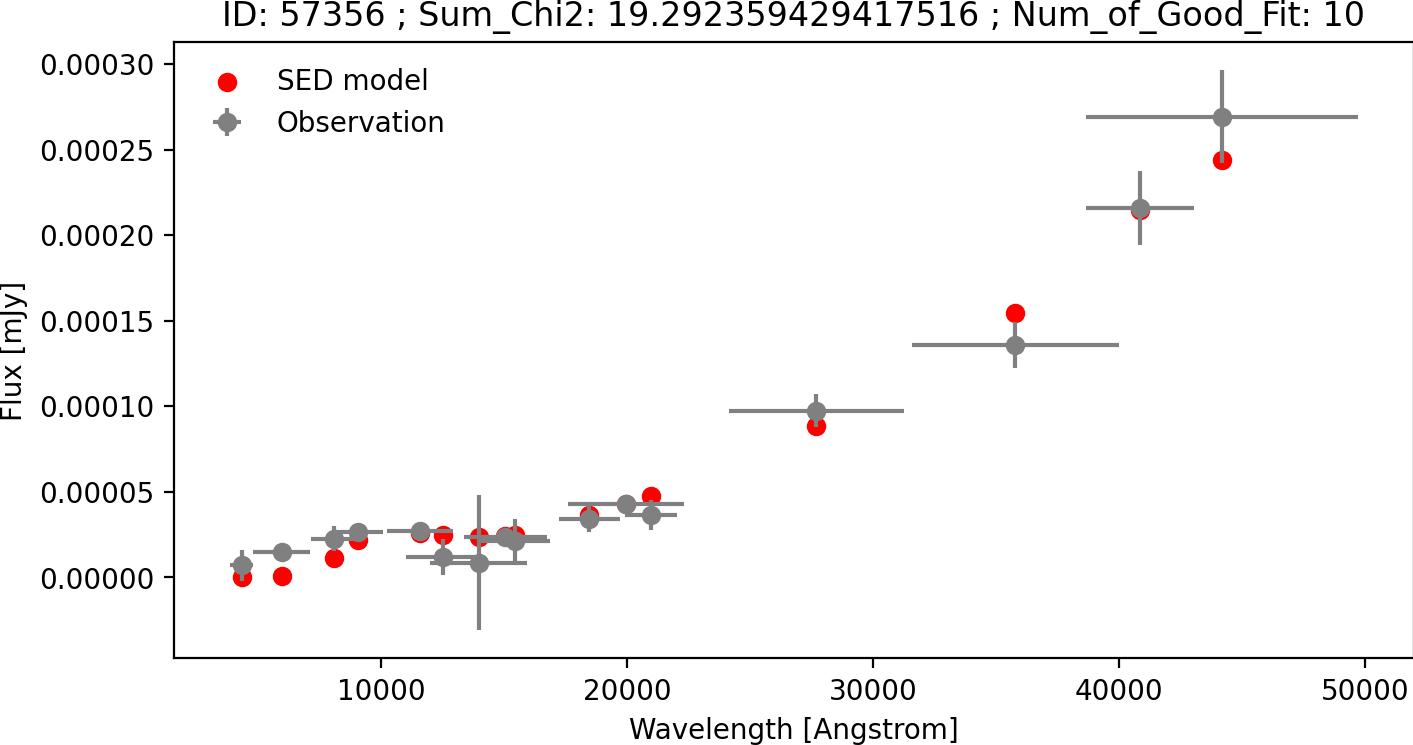}}
    \subfigure[$A_v=0.36\pm0.48\,\mathrm{mag}$ ; $\log M_*/M_{\odot}=9.88\pm0.08$ ; $\mathrm{frac}_{\mathrm{AGN}}=0$ ; $z_{phot}=5.5$]{\includegraphics[width=0.68\linewidth]{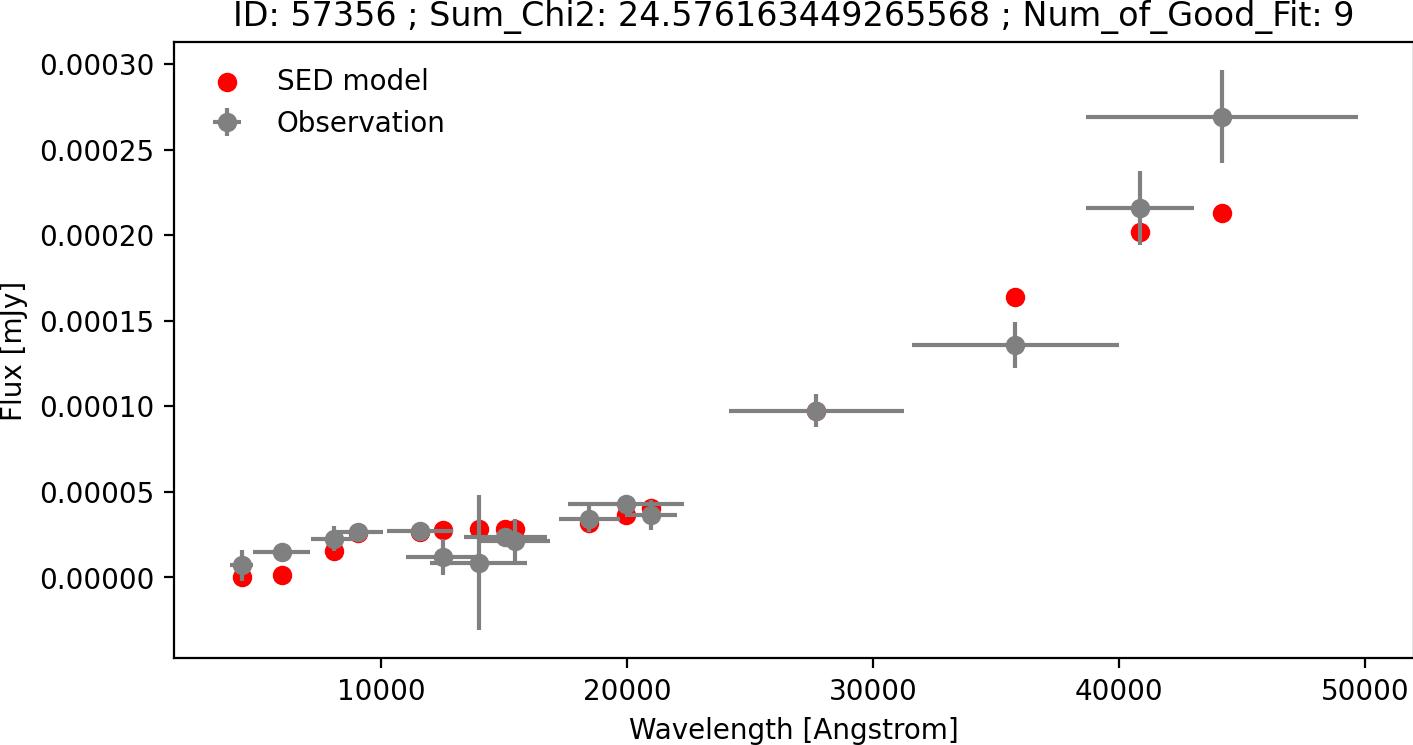}}
    \caption{SED Fitting Results for JADES LRD 57356. Panel (a) shows a dustier solution with a dominant AGN, while panel (b) presents a less dusty, purely star-forming galaxy.}
    \label{fig:lrd_57356}
\end{figure}

\clearpage 

\bibliography{main}
\bibliographystyle{neurips_2024}
\end{document}